\documentclass[acmsmall]{acmart}

\AtBeginDocument{%
  \providecommand\BibTeX{{%
    \normalfont B\kern-0.5em{\scshape i\kern-0.25em b}\kern-0.8em\TeX}}}

\setcopyright{acmcopyright}
\copyrightyear{2018}
\acmYear{2018}
\acmDOI{10.1145/1122445.1122456}

\acmJournal{JACM}
\acmVolume{37}
\acmNumber{4}
\acmArticle{111}
\acmMonth{8}

\usepackage{lmodern} 

\usepackage{booktabs}
\usepackage{algorithm}
\usepackage[noend]{algpseudocode}
\usepackage{subcaption} 
\usepackage{multirow}
\usepackage{tabu}
\usepackage{float} 

\begin{document}

\title{Fortifying Vehicular Security Through Low Overhead Physically Unclonable Functions}

\author{Carson~Labrado}
\author{Himanshu~Thapliyal}
\email{hthapliyal@uky.edu}
\affiliation{%
  \institution{University of Kentucky}
  \department{Department of Electrical and Computer Engineering}
  \city{Lexington}
  \state{Kentucky}
  \postcode{40506}
  \country{USA}
}

\author{Saraju~P.~Mohanty}
\email{saraju.mohanty@unt.edu}
\affiliation{%
  \institution{University of North Texas}
  \department{Department of Computer Science and Engineering}
  \city{Denton}
  \state{Texas}
  \postcode{76207}
  \country{USA}
}

\renewcommand{\shortauthors}{Labrado, et al.}

\begin{abstract}
Within vehicles, the Controller Area Network (CAN) allows efficient communication between the electronic control units (ECUs) responsible for controlling the various subsystems.  The CAN protocol was not designed to include much support for secure communication.  The fact that so many critical systems can be accessed through an insecure communication network presents a major security concern.  Adding security features to CAN is difficult due to the limited resources available to the individual ECUs and the costs that would be associated with adding the necessary hardware to support any additional security operations without overly degrading the performance of standard communication.  Replacing the protocol is another option, but it is subject to many of the same problems.  The lack of security becomes even more concerning as vehicles continue to adopt smart features.  Smart vehicles have a multitude of communication interfaces would an attacker could exploit to gain access to the networks.  In this work we propose a security framework that is based on physically unclonable functions (PUFs) and lightweight cryptography (LWC). The framework does not require any modification to the standard CAN protocol while also minimizing the amount of additional message overhead required for its operation.  The improvements in our proposed framework results in major reduction in the number of CAN frames that must be sent during operation.  For a system with 20 ECUs for example, our proposed framework only requires 6.5\% of the number of CAN frames that is required by the existing approach to successfully authenticate every ECU.  
\end{abstract}



\maketitle

\section{Introduction}
Vehicles are no longer a purely mechanical machine.  Modern vehicles include a not insignificant number of digital components such as infotainment systems and the electronic control units (ECUs) that are responsible for controlling the various subsystems within the vehicle.  These devices are connected via various intra-vehicle networks, the most notable being the Controller Area Network (CAN) which provides a relatively inexpensive method for several ECUs to communicate with each other \cite{Davis2007}.

Unfortunately, CAN was not developed with security in mind.  The lack of security has become much more alarming over the last decade as researchers have been able to successfully attack vehicles by exploiting the shortcomings in the CAN protocol.  A very notable example of this was in 2015 when researchers were able to control a consumer vehicle \cite{miller2015remote}.  Due to the nature of CAN, the ECUs for such systems as the engine, brakes, and steering were all connected to the same CAN bus.  All an attacker needs to do to carry out an attack is gain access to the CAN bus.  This could be achieved through somehow compromising an ECU or more simply creating their own connection.  The lack of security features means that all transmitted messages are treated as being from a valid source regardless of their actual origin.  For example, an attacker could send a message instructing the vehicle to apply the brakes.  The vehicle would comply as it has no way of verifying the validity of the message.

The issue of vehicular security is likely to become even more pressing in the coming years.  This in large part can be contributed to the continual push to develop fully autonomous vehicles in addition to the inclusion of smart features in vehicles.  Every new type of connection added to a vehicle represents a new potential attack surface for malicious actors.  Some of the connections include vehicle-to-vehicle (V2V), vehicle-to-network (V2N), vehicle-to-infrastructure (V2I), and vehicle-to-pedestrian (V2P). This overall connected environment is collectively referred to as vehicle-to-everything (V2X) \cite{chen2017vehicle}.  

\begin{figure}[h]
\centering
\includegraphics[width=4in]{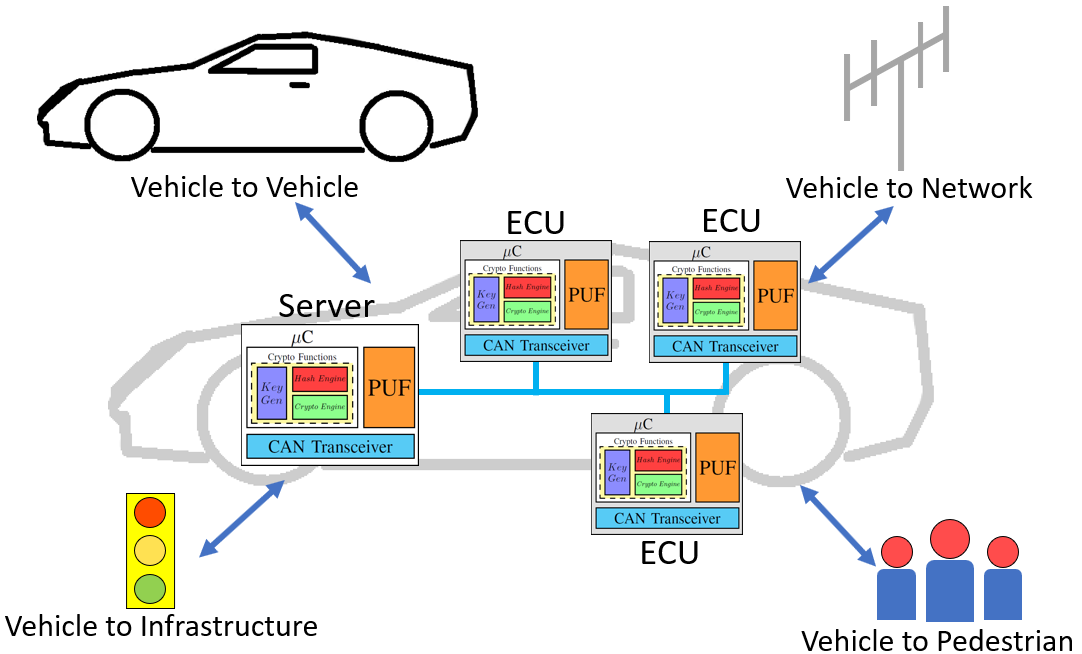}
\caption{PUF Integrated Smart Vehicle}
\Description{Shows a vehicle with an integrated PUF security framework.  Connections are shown between the vehicle and other vehicles, infrastructure, pedestrians, and external networks.}
\label{V2X}
\end{figure}

It is difficult to design a singular security solution since a vehicle's expanded features are provided by separate subsystems which communicate via in-vehicle communication networks.  There in fact exists a knowledge gap in terms of how damage could by caused by attacking various components and systems of the vehicle.  For example, it has been shown to be possible to compromise some of a vehicle's sensors in order to trick the driver and/or the vehicle's control system.  However, it is unknown just how vulnerable all of the sensors are and how severe of a reaction can be induced by a hacked sensor generating erroneous readings \cite{parkinson2017cyber}. 

This is a constantly evolving problem that demands regularly devising new techniques to combat previously unknown vulnerabilities.  However, vulnerabilities that have been known for years demand a similar level of time and focus.  A major target should be devising solutions for the inherent vulnerabilities in the CAN bus.  Because CAN is such a fundamental communication network, any potential security solutions should seek to remain as true as possible to the original specification.  Major deviations could result in the need to redesign an untold number of internal systems to make them compatible with the new solution.

In this work we propose a new security framework that adds security features while minimizing overhead and without making any changes to the basic CAN protocol.  This framework is a server-based approach where a central server connected to the CAN bus is responsible for authenticating all connected nodes and generating session keys.  The design utilizes physically unclonable functions (PUFs) as the basis for key storage, key generation, and the authentication of the nodes. Lightweight cryptographic algorithms are employed as they are more aptly suited than standard cryptographic algorithms to the resource constrained environments of vehicles.  Figure \ref{V2X} shows an example of the proposed framework incorporated into a smart vehicle environment.

The rest of this work is organized as follows: Section II presents our vision for using PUFs as a low overhead smart car security solution; Section III describes related work on consumer electronics security and provides background information on CAN, its vulnerabilities, and PUFs; Section IV describes the proposed framework's operation in detail; Section V analyzes the framework and its security capabilities; Section VI provides a comparison to other PUF-based security frameworks; lastly, Section VII concludes the paper.

\section{Our Vision for PUF-based Low Overhead Smart Car Security}
We believe that the security challenges facing vehicles are so unique that classical security approaches alone will not be sufficient.  Vehicles are designed such that some of their core functionality is directly provided by inherently insecure components and subsystems.   These components are in fact so well ingrained that replacing all of them would likely require vehicles to be fundamentally redesigned from the ground up.  While an approach like this could work in theory, the sheer cost of design, not to mention the material costs of the new components, would seem to prevent this from being a truly viable option for anything short of very long term goals.  

In addition to providing security, we believe that for a security solution to be more immediately viable it should minimize both the monetary and computational costs that would be incurred by its introduction to the vehicle.  As such, there are three major design goals that vehicle security solutions should strive to meet: 

\begin{enumerate}
\item Minimize additional hardware and computation.
\item Avoid significant changes in protocols.
\item Minimize communication overhead.
\end{enumerate}

First, security solutions should seek to minimize the addition of extra hardware and computation.  The resource constrained and real-time nature of vehicles does not allow for much extra computation for things like encryption.  Upgrading the existing devices or adding specialized hardware to provide the resources needed for the additional computation would drive up implementation costs. 

Second, solutions should not make any significant changes to protocols.  We consider significant changes to include any modification that would require likewise changes in the supporting hardware.  For example, replacing a communication protocol or adding additional message fields cannot occur without upgrading the current infrastructure to support the new features.  

Lastly, a solution should try to minimize communication overhead so that it does not effectively violate the second goal without actually changing the protocol itself.  Some protocols like CAN can only send a very limited amount of data per message.  Breaking a single transmission across multiple CAN messages allows for the transmission of cryptographic keys and encrypted data, but at the cost of reducing the bandwidth and responsiveness of the network.

\begin{figure}[h]
\centering
\includegraphics[width=3.2in]{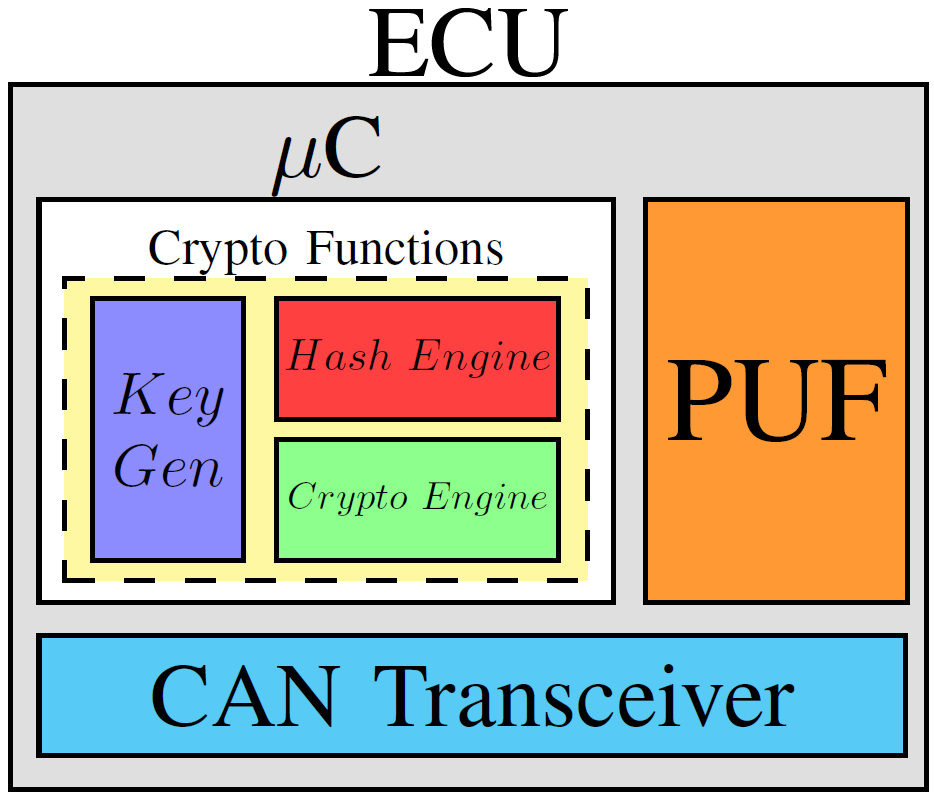}
\caption{ECU PUF Integration}
\label{PUFECU}
\Description{Shows example PUF ECU integration.  Consists of a PUF, CAN transceiver and microcontroller.  The microcontroller contains support for cryptographic functions such as key generation, hashing, and encryption.}
\end{figure}

Many potential security solutions would introduce additional overhead in both computation and the number of additional messages that must be sent in support of the normal transmission of data \cite{wu2016security} \cite{king2017international} \cite{siddiqui2017secure}.  Furthermore, some solutions would likely require the addition of hardware for features such as secure key storage and generation, data encryption, etc.  The use of physically unclonable functions (PUFs) in security solutions could potentially provide a cheaper option for the implementation of some of the these features. PUF-based security solutions would thus more closely align with our previously stated design goals.  It is for that reason that our proposed security framework is directly based on PUFs.  An example of this integration, which is used by our proposed framework, is shown in Figure \ref{PUFECU}.  Every ECU would include its own PUF which could then be utilized by a variety of security operations.  Integration of a PUF in this manner would leave open the possibility of maintaining the underlying CAN protocol and thus should not require any modifications to the actual CAN network infrastructure that connects the ECUs.

\section{Background}
\subsection{Prior Related Work on Consumer Electronics Security}
Providing security to vehicles is a challenging problem that is not readily solvable by conventional solutions.  Even just the diagnosis of security threats has required the development of novel intrusion detection methods \cite{moore2019data}.  However, the security issues facing vehicles are not as isolated as they might appear.  Similar security concerns are actually being raised in a variety of other areas.  This general trend is a direct response to society's adoption of the Internet of Things (IoT).  The addition of smart features to an increasing number of consumer electronics has also introduced security vulnerabilities and concerns that were not present when the devices were originally designed.

Developing methods to combat these new challenges has drawn interest from a number of researchers.  This has included classical approaches such as designing hardware security chips for mobile devices \cite{ju2015implementation} and secure firmware validation and update schemes for personal home devices \cite{choi2016secure}. Other areas of interest include security architecture for edge devices \cite{tiburski2019lightweight} and protecting the runtime data of embedded systems through hardware-enhanced cryptographic engines including AES and the hashing algorithm LHash \cite{wang2019hardware}.

Researchers have also taken to examing more novel security approaches such as creating PUFs that are specifically designed for use in IoT applications.  This has included both adaptations of established designs such as Ring Oscillator (RO) PUF \cite{khan2019ultra} along with novel approaches such as designs based on adiabatic logic \cite{kumar2020design} and bloom filters on memristor-based PUFs \cite{lee2020novel}.  Researchers have explored how PUFs such as these could serve as the basis for more complete security frameworks and systems.  One interesting example is a framework in which individual embedded devices use PUFs to create their own unique fingerprints \cite{huang2020puf}.  Those fingerprints are then encoded in order construct a larger system-level fingerprint. In this way the system level ID can be used to identify if one of the system's individual devices is no longer valid.  Another approach utilizes memristor-based PUFs to create a very lightweight security system \cite{uddin2019memristor}.  The PUFs operate as a one time pad by generating a random key each time one is needed for an encryption and decryption operation.  A random response is sent to the PUF and the associated response is used as the key.  Other research efforts have included using PUFs to create novel device authentication schemes for IoT-enabled medical devices \cite{yanambaka2019pmsec} and radio-frequency (RF) communication between nodes in a wireless network \cite{chatterjee2019rfpuf}.  The inclusion of PUFs has the potential to thus introduce security features into an intra-vehicle network while minimizing any changes in its normal operation.

\subsection{Controller Area Network (CAN)}
\begin{figure}[h]
\centering
\includegraphics[width=4.5in]{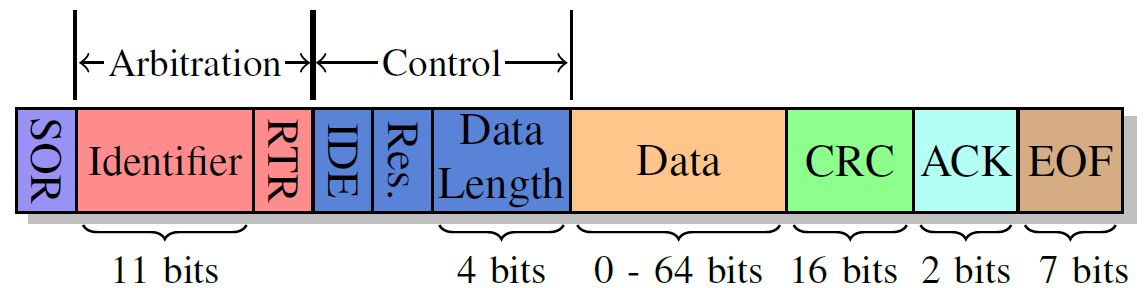}
\caption{CAN Frame}
\label{CAN}
\Description{Shows the various bit fields within a standard CAN frame.  There are fields for arbitration, control, data, CRC, ACK, and EOF.  Arbitration contains an 11 bit identifies and control contains a 4 bit data length field.  Data is 0-64 bits, CRC is 16 bits, ACK is 2 bits, and EOF is 7 bits.}
\end{figure}

The Controller Area Network (CAN) is a serial communication system that allows for simple and efficient message passing between connected nodes without requiring a master controller in the network \cite{Davis2007}.  CAN is most commonly used in vehicles to allow communication between the embedded electronic control units (ECUs) without having to implement point to point wiring between all possible communication paths.  Figure \ref{CAN} shows the format of a standard CAN frame or message.  A standard CAN frame has a very limited number of message fields.  The arbitration portion denotes the ID of the message.  The control field shows the number of bytes of data (0-8 bytes) being sent by the frame.  CRC stands for cyclic redundancy check and is an error correcting code used to check for errors in the transmission.  ACK is used to denote if a message was successfully received.  Lastly, EOF denotes the end of the frame.

\subsection{CAN Vulnerabilities}
The CAN protocol was not originally designed to include much in the way of security features.  The key issues are messages are broadcast to all connected nodes, the data fields are not encrypted, and there is no way to authenticate or even known who was responsible for sending a given message.  An attacker only has to gain access to the CAN bus in order to carry out a wide arrange of attacks including eavesdropping, spoofing/impersonation, and denial of service (DoS).  Through eavesdropping an attacker would be able to monitor all communications and launch a replay attack by sending a duplicate of a previously seen message \cite{koscher2010experimental}.  Another possibility would be to reverse engineer what would be the (likely manufacturer-specific) communication protocol used between nodes.  Once that was accomplished, an attacker would be able to send erroneous messages that the targeted ECUs would interpret to be valid due to CAN's inherent lack of authenticity.  Researchers have shown that attacks of this nature can be utilized to control different components of the vehicle such as controlling the dashboard and shutting off the engine \cite{woo2015practical}.

The CAN protocol also makes CAN very susceptible to DoS attacks.  The CAN standard guarantees that the message with the highest priority will be the first message to go through.  If the CAN bus is currently in the process of transmitting a message, it will stop that transmission and begin to transmit the new message provided that new message has a higher priority.  An attacker only has to repeatedly transmit high priority messages for the CAN protocol to guarantee that the messages from ECUs will never get a chance to send due to having a lower priority \cite{buttigieg2017security}.

\subsection{Physically Unclonable Functions}
Physically Unclonable Functions (PUFs) are a class of device that utilize internal variations introduced by the manufacturing process to generate unique outputs for a given input.  The input to a PUF is denoted as a ``challenge'' and the output is known as a ``response''.  A challenge and its associated response are collectively known as a challenge-response pair (CRP).  For a given challenge, the response produced by different PUFs should be unique since each response is a direct manifestation of the unique physical properties of that specific PUF.  Furthermore, a PUF with a small number of CRPs, typically just one, is a weak PUF and a PUF with a large number of CRPs is considered to be a strong PUF.

PUF designs are commonly based on transistor level process variations such as gate delays \cite{hori2010quantitative} or the initial power-on value in memory cells \cite{merli2010improving}.  Other researchers have explored creating PUFs from larger components such as energy harvesters \cite{nozaki2019energy} and sensors \cite{labrado2019use}.  The unique properties of PUFs make them an intriguing option as a low cost method for implementing security related features such as key storage \cite{feiri2013efficient} or hardware obfuscation \cite{wendt2014hardware}.

\section{Proposed CAN Security Framework}
The overall design of our proposed framework involves using a server within the network to authenticate all nodes before allowing normal message passing operations to begin. The proposed framework requires an LWC functions for encryption, decryption, and hashing.  Any LWC function can be used as long as it meets certain criteria.  The LWC function used for encryption and decryption must have a block size of 64 bits and a key size of no more than 128-bits.  For the LWC hash function, it must be able to generate 128-bit hashes. 

As an example, our proposed framework is described in terms of using PRESENT \cite{bogdanov2007present} for encryption and PHOTON \cite{guo2011photon} for hashing.  We use these as examples as they have both been defined as International Organization for Standardization (ISO) standards for LWC\cite{isoblock} \cite{isohash}.  Either LWC function could however be substituted with a different one which meets the aforementioned criteria.  The proposed framework also makes use of Elliptic Curve Diffie-Hellman (ECDH) key exchange based on FourQ which has been shown to offer better performance than other curves targeting the same level of security \cite{costello2015fourq}.  These cryptographic algorithms will be discussed in more detail in Section \ref{ANALYSIS}.  The proposed framework supports 80-bit or 128-bit encryption keys.  For the sake of simplicity, the figures and tables in this section assume 80-bit encryption keys.

Our proposed framework is not designed for use with only one specific PUF design.  It is assumed that the chosen PUF will be a weak PUF since the framework needs a given PUF to always produce the same response each session.  The keys are derived from the PUF responses so the keys would change if the response changed.  A strong PUF could be an option if it was configured to operate as a weak PUF by always providing it the same response.  Topics related to the actual implementation of the PUF should be considered outside the scope of this paper. This includes methods for improving the reliabilities of PUFs such as error correcting codes and other schemes.  Additional resources required for a specific PUF implementation are likewise a direct result of the chosen PUF rather than our proposed framework.

The proposed framework can be divided into the distinct operation phases of enrollment, authentication, and normal operation.  The authentication and normal operation phases will occur every time the system is turned on.  By contrast, enrollment would ideally only ever occur once for the entire existence of the system.  The rest of this section describes each of the phases in greater detail.

\subsection{Enrollment}
This phase should only occur once, likely during the manufacturing phase.  This should in theory provide a secure environment for data to be hardcoded into the server and other nodes.  The purpose of the enrollment phase is to give each node a copy of the server's public key.  This allows each node to ultimately derive a shared secret with the server that it can use to securely communicate with the server during the authentication phase.  The server will likewise need to have a copy of the public key for every node.  In addition, the server needs to store a hash of the response from each node.  The response hashes are 128 bits in size which means future stages will only need 2 CAN frames to transmit the entire hash.

We use response hashes rather than raw responses for two main reason.  The first reason is it allows greater flexibility in choosing a type of PUF to use within the framework.  Choosing a PUF with a response larger than 128 bits won't increase the number of CAN frames required for a node to send it to the server.  The other reason is this prevents sending the PUF's response outside of the node.  Even though the responses would be encrypted, the server would still need to do a comparison with the unencrypted response in order to validate it.  The raw response can be used to directly generate a secret key, while it is not possible to do the same with a hash of the response.  This removes the need to take the same security precautions with storing and handling the response hash that you would need if you were instead using the raw response.

\begin{figure}
\centering
\includegraphics[width=4in]{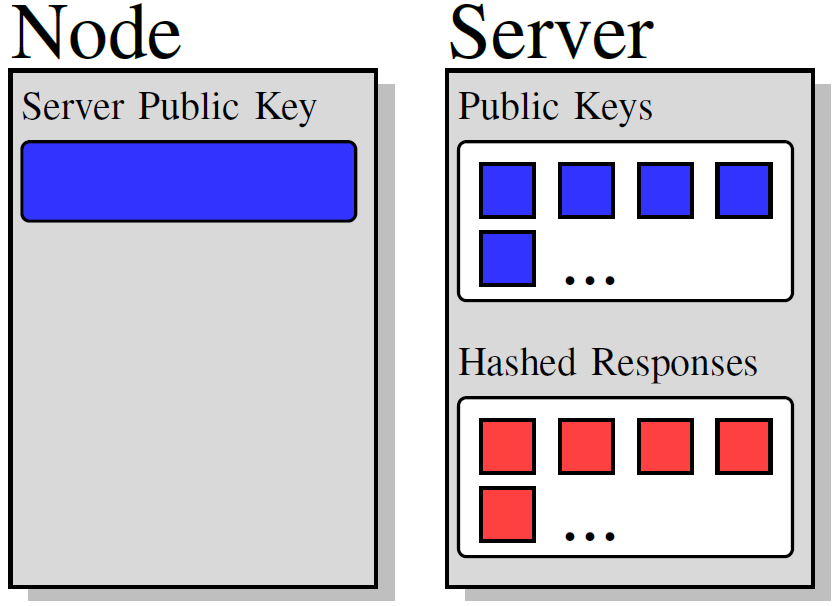}
\caption{Post-Enrollment Stored Values}
\label{Enrollment}
\Description{A node will only store a copy of the server's public key.  The server will have stored copies of multiple node public keys and node hashed responses.}
\end{figure}

During authentication a node will be considered valid if it is able to generate a response whose hash matches the associated one stored by the server.  Figure \ref{Enrollment} provides a visualization of what data will be stored within each entity at the end of the enrollment phase.

\subsection{Authentication}
\begin{figure}
\centering
\includegraphics[width=3in]{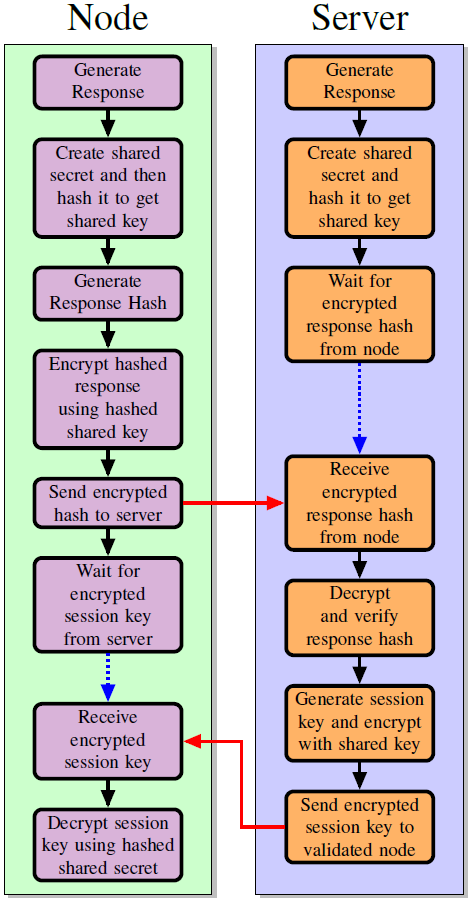}
\caption{Authentication Process}
\label{Authenticate}
\Description{Illustrated example of the steps mentioned in Algorithms 1 and 2.}
\end{figure}

The authentication phase should run every time the network is first powered on.  Within this phase the server will first validate the authenticity of each node in the system. Next it will generate a session key and send a hashed copy to each node to use during normal operation. Algorithm \ref{NodeAuth} describes the individual steps taken by a given node and Algorithm \ref{ServerAuth} describes the steps for the server. These steps assume 80-bit encryption keys.  In addition, the entire authentication process is illustrated in Figure \ref{Authenticate}.


\begin{algorithm}
\caption{Node Authentication Process}
\label{NodeAuth}
\begin{algorithmic}[1]
	\State The node's PUF generates a response $\mathbf{R}$.
	\State A 128-bit hash of the response is created $\mathbf{H_R}$.  The response is also used as the node's secret key $\mathbf{x}$ by FourQ.
    \State A shared secret $\mathbf{SSec}$ between the node and the server is generated using the node's private key $\mathbf{x}$ and the stored public key of the server $\mathbf{P_S}$.
    \State The shared secret is hashed and truncated to produce an 80-bit key $\mathbf{K_{SSec}}$.
    \State The node's hashed response $\mathbf{H_R}$ is encrypted using the hashed shared secret as the key $\mathbf{K_{SSec}}$.
    \State The encrypted response hash is sent to the server.  
    \State The node then waits for the server to respond with an encrypted session key.
    \State The node decrypts the session key $\mathbf{K_{Sess}}$ using the hashed shared secret as the key $\mathbf{K_{SSec}}$.
    \State The list of valid nodes is extracted from the decrypted session key if the system was configured to support it.
    \State This session key $\mathbf{K_{Sess}}$ will later be used during normal operation to encrypt and decrypt all messages within the network.
\end{algorithmic}
\end{algorithm}

\begin{algorithm}
\caption{Server Authentication Process}
\label{ServerAuth}
\begin{algorithmic}[1]
        \State The server's PUF generates a response $\mathbf{R_S}$.
        \State The response is used as the server's secret key $\mathbf{x_S}$ by FourQ.
        \State A shared secret $\mathbf{SSec}$ between a given node and the server is generated using the server's private key $\mathbf{x_S}$ and the stored public key of the node $\mathbf{P}$.
        \State The shared secret $\mathbf{SSec}$ is hashed and truncated to produce an 80-bit key $\mathbf{K_{SSec}}$.
		\State The server waits to receive encrypted response hashes from each node.
		\State The server decrypts the hashes using the hashed shared secret associated with that specific node as the key $\mathbf{K_{SSec}}$.
		\State The decrypted response hashes $\mathbf{H_R}$ are validated by comparing them to previously stored hashes.
		\State The server generates a random session key $\mathbf{K_{Sess}}$ and concatenates it with either padding or a bit mask representing valid nodes in the network.
		\State The server encrypts a copy of the concatenated session key $\mathbf{K_{Sess}}$ for each node using the hashed shared secret associated with that node as the key $\mathbf{K_{SSec}}$.    
		\State The server sends an encrypted session key to each node. 
\end{algorithmic}
\end{algorithm}

It is important to note that the shared secret between each node and the server is not directly used as a key for encryption and decryption.  We hash the shared secret to get a shared key.  This helps prevent key leakage and reduces the shared secret to the key size required by the encryption algorithm.  The nodes use the shared key to transmit their response hashes and the server uses it to transmit the session keys.

\begin{figure}
\centering
\begin{subfigure}{0.6\textwidth}
\centering
\includegraphics[width=3in]{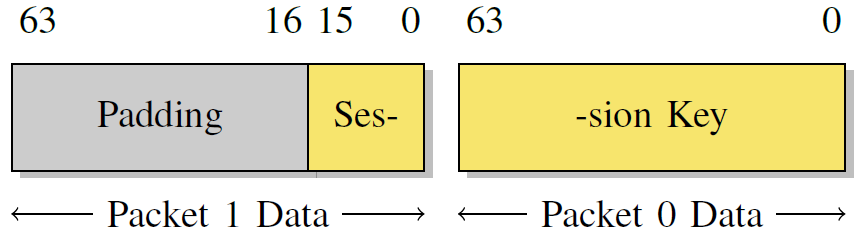}
\caption{80-bit Session Key}
\label{fig:padding}
\Description{Two 64 bit data packets.  The session key occupies 80 bits and the other 48 bits is padding}
\end{subfigure}

\begin{subfigure}{0.6\textwidth}
\centering
\includegraphics[width=3in]{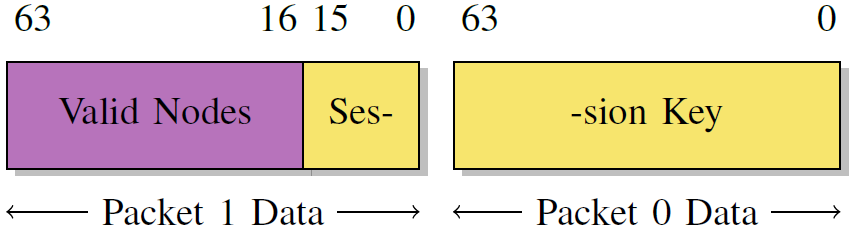}
\caption{48-bit Valid Bitmask and 80-bit Session Key}
\label{fig:mask}
\Description{Two 64 bit data packets.  The session key occupies 80 bits and the other 48 bits is used to denote which nodes are valid.}
\end{subfigure}

\begin{subfigure}{0.6\textwidth}
\centering
\includegraphics[width=3in]{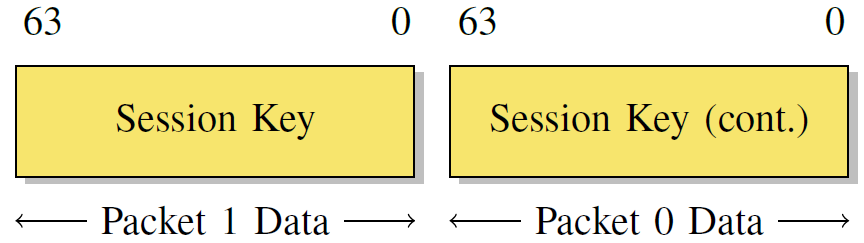}
\caption{128-bit Session Key}
\label{fig:full}
\Description{Two 64 bit data packets.  The session key occupies all 128 bits.}
\end{subfigure}

\caption{Options for Encrypted Session Key Packets}
\label{AUTH_PACKETS}
\Description{Three options are presented.  See the individual subfigure descriptions for information about each one.}
\end{figure}

The only messages sent during this phase are the encrypted response hashes and the encrypted session keys.  The encrypted response hashes are 128 bits in size which means it will take 2 CAN frames to transmit the entire hash.   Similarly, the encrypted session key must be a multiple of 64 bits in order to minimize the amount of CAN frames required to transmit it.  This allows us to support key sizes of 80 and 128 bits.  80-bit keys would need to be concatenated with 48 bits of padding.  Alternatively, an 80-bit key could be concatenated with a 48-bit wide bitmask that denotes the nodes that were successfully Authenticated.  Each bit would correspond to a specific node.  Figure \ref{AUTH_PACKETS} provides an illustration of these different modes of operation.  

\subsection{Normal Operation}
\begin{figure}
\centering
\includegraphics[width=3.2in]{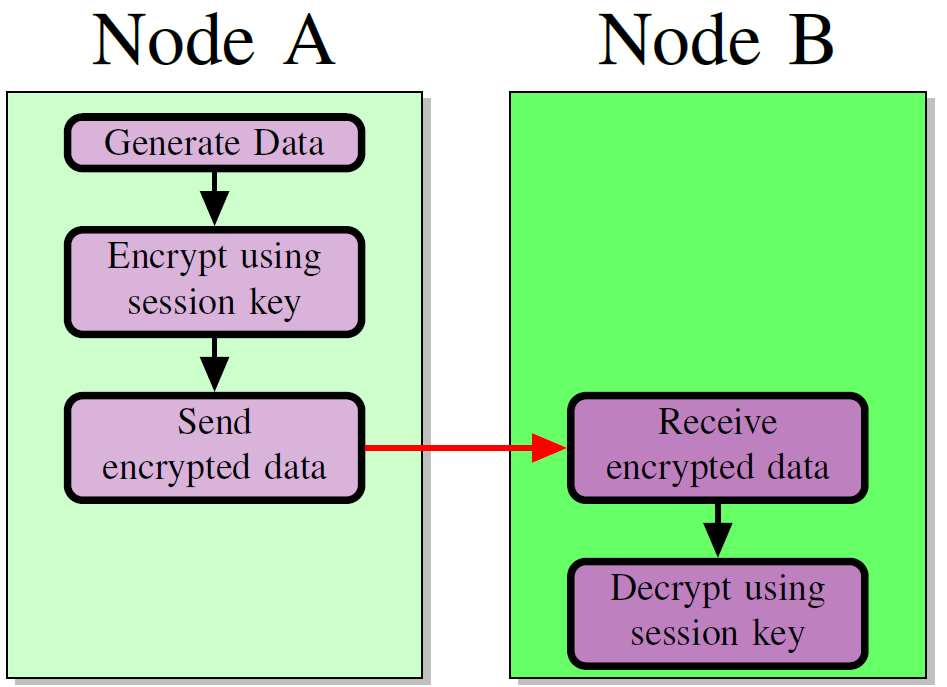}
\caption{Normal Communication Between Two Nodes}
\label{NORMAL}
\Description{Node A generates data, encrypts it using the session key, and then sends the data to Node B.  Node B receives the encrypted data and then decrypts it using the session key.}
\end{figure}

This phase is analogous to the way a normal CAN bus operates and ultimately serves the same purpose.  The major difference is all transmitted data must be encrypted before it is sent over the bus.  Once nodes have have been authenticated by the server and received a session key, the system can transition to normal communication between nodes in the network. The session key is used to encrypt the data field of a packet before sending it across the CAN bus to another node.  That other node can then use its own copy of the session key to decrypt the data and then respond accordingly.  Figure \ref{NORMAL} shows the general flow for one node communicating with another node.

\section{Design and Analysis of Proposed Framework}
\label{ANALYSIS}
By deriving keys from physically unclonable functions (PUFs), we avoid the need to use costly secure nonvolatile memory for key storage.  Instead, the keys can be generated as needed during each authentication phase.  This means an attacker would have to obtain physical access to the PUF to recover its response and associated key pair.  As we will explain, the information that must be persistently stored between sessions does not necessarily need to be kept secret and that allows us to save costs by not requiring secure nonvolatile memory in the nodes.

The public keys can be stored in unsecured memory since they require a separate private key to form a shared secret.  The private key is generated whenever it is needed and deriving a private key from its associated public key would require successfully breaking the elliptic curve key generation cryptographic algorithm.  The hashed responses can also be stored in unsecured memory since the server expects any received hashed responses to be encrypted with the appropriate shared secret that it is assumed an attacker is not able to obtain.  Furthermore, since hash functions are considered to be one-way it should not be possible to recover the original input response that produced the hash and then use that response to generate its associated private and public keys.

During each new session, the server generates a session key that will be used by all ECUs during normal communication.  Depending on the mode of operation, that key can be concatenated with a bit mask denoting which nodes are valid.  During authentication a server with $n$ ECUs will receive $n$ hashed responses and then only have to transmit $n$ total copies of the generated session key, one for each ECU.  Therefore, the number of frames that must be sent in our proposed framework scales linearly as the number of ECUs in the system increases.  Furthermore, the hashed responses and session key payloads can each be transmitted in only 2 frames.  This means that during the authentication phase of our proposed framework, a network with $n$ ECUs will require the transmission of $4n$ total frames to complete the authentication phase. 

The overhead in terms of frames required by the proposed framework is shown in Table \ref{COMPTAB}.  The enrollment phase is omitted since it would likely not require sending any messages over the CAN bus and would ideally only ever run once.  A partial repeat of the enrollment phase would only need to occur when PUFs are added and/or removed from the system, e.g. completely replacing a malfunctioning node.  The enrollment phase is otherwise completely implementation dependent and occurs outside of the flow of operations for the system.  During normal operation, our proposed framework operates exactly the same as the standard CAN protocol.  The same number of messages are required to transmit the same amount of data.  The only real difference is the data contained within the data field of the message is now encrypted.  The cryptographic operations will of course introduce some additional overhead, but that will be highly dependent upon the chosen algorithms and the underlying hardware.  Special purpose hardware for example could greatly speed up calculations or certain cryptosystems may perform better on the specific ECUs used by a given manufacturer.

\subsection{Threat Mitigation}
The major threat that will be directly mitigated is eavesdropping.  Currently an attacker with access to the CAN bus can see all messages that are transmitted.  Our proposed framework counteracts this by encrypting the actual data that is transmitted.  The only potentially useful information that could then be used by an attacker is the destination IDs of the messages.  

Other notable attacks are data tampering and impersonation attacks.  As the name suggests, data tampering attacks occur when an attacker is able to successfully modify a message without the sender or receiver being able to detect that the message has been changed.  Impersonation attacks are where an attacker impersonates another ECU and sends messages as if they were that ECU.  These attacks are especially concerning because they can allow an attacker to effectively control a vehicle.  The CAN protocol has no built in mechanism for identifying the original sender of any message.  An attacker for example could send messages to engage the brakes and the brakes would activate as if the associated ECUs had received legitimate commands.  Our proposed framework also provides some protection against these sorts of attacks.  The first step of being able to forge messages is to understand the actual message format.  Doing so requires an attacker to reverse engineer the message protocol by monitoring the network.  If the actual message format is not already known by an attacker, then it will be difficult for them to reverse engineer it since the data itself will be encrypted within our framework.  

In the event an attacker does know the message format, it will still be difficult for them to create erroneous messages to produce specific outcomes like applying the brakes.  All messages are encrypted with a session key so an attacker would need to have a copy of that key in order to properly encrypt their message.  Otherwise, their message will get mangled when the receiving node attempts to decrypt it.  This should force the attacker to resort to a replay attack in which they capture a message and then repeat it to produce a known result.  This is much more time consuming since the attacker would have to try to monitor the entire network traffic and somehow correlate a specific message payload to a specific ID with producing a desired response in the vehicle.  Since the session key is randomly generated each time, the encrypted form of a given message will change each time the session key changes.  The attacker would then have to repeat the entire process every time a new session begins.  This prevents an attacker from simply building a library of messages across several sessions since the messages will change each time.

The types of security threats that our framework does not offer much protection against are those that do not require reading and/or writing specific data values.  Attacks of this nature succeed by merely transmitting a message regardless of its actual content.  One notable example of this type of attack is a Denial of Service (DoS) attack.  A DoS attack would seek to disable a vehicle by flooding the CAN bus with high priority messages.  The higher priority means that these erroneous messages will get delivered before the valid, yet lower priority messages required for normal operation.  The valid messages never get delivered and the vehicle is thus unable to function.  Certain forms of data tampering and impersonation attacks would also fall under this type of attack.  The goal of these attacks would not be producing a specific outcome such as controlling the vehicle's movement.  Instead, they would seek to cause general havoc by either repeating previously seen packets or sending what would amount to junk data to ECUs. The attacker would have no notion of what the outcome will be.  It would instead be completely up to chance in terms of how the vehicle will actually respond.  As such, the possible response could range in severity from effectively ignoring the attacker's messages, all of the way to actually causing some sort of accident.

\subsection{Cryptographic Algorithms}
The performance of current cryptographic standards is not always suitable for use in resource constrained environments.  Lightweight cryptography (LWC) seeks to address this by specifically designing cryptographic algorithms for resource constrained environments \cite{nist2019lightweight}. Although the National Institute of Standards and Technology (NIST) is currently in the process of setting LWC standards, the International Organization for Standardization (ISO) has published LWC standards. 
 
Our adherence to lightweight cryptographic algorithms \cite{isoblock} \cite{isohash} should provide other performance benefits and potentially reduce the cost of implementation.  The amount of computation required to perform basic cryptographic operations such as encryption, key generation, etc., is reduced in our proposed framework compared to existing solutions which utilize larger algorithms such as AES.  This has the added benefit of potentially simplifying any dedicated hardware that is solely designed to perform those operations.  Furthermore, our design does not require any form of secure nonvolatile memory for key storage as the use of a PUF allows all keys to be generated as needed.

\subsubsection{ECDH based on FourQ}
For key exchange we used Elliptic Curve Diffie-Hellman (ECDH) based on FourQ.  FourQ is an elliptic curve which targets the 128-bit security level \cite{costello2015fourq}.  Although other curves targeting the 128-bit security level would also work, FourQ has been shown to be faster than other popular 128-bit security elliptic curves such as NIST P-256 and Curve25519 in both key generation and secret exchange \cite{costello2015fourq} \cite{alvarez2017algorithms}. 

\subsubsection{Encryption and Decryption}
The use of a block cipher which has a block size of 64 bits means that during regular communication, the number of CAN frames that must be used to send encrypted data will remain the same as the number that must be used to send normal unencrypted data.  Any additional overhead introduced by the proposed framework during normal operation would thus be solely limited to the encryption and decryption operations performed by each node.  

\subsubsection{Hash Function}
It was important to choose a hash function that produced 128-bit hashes as it would allow us to use hashes as encryption keys.  The other major benefit is hashed responses can be sent in just 2 CAN frames.  
 
\subsection{Server Capabilities}
It is assumed that the server will be secure and have the capability to securely generate a random session key for each session.  It should not be possible to add, remove, and/or modify nodes and the public keys and hashed responses associated with them except during the enrollment phase in a trusted environment.  The server also has the potential to act as a monitor of sorts during the authentication phase.  It could phase certain anomalies as malicious and either lock out the rest of the authentication phase, or notify a more centralized security system so that it may act accordingly.  The CAN protocol does not show the origin of messages being transmitted.  However, it can still detect situations such as multiple authentication attempts for a single node, an incorrect number of nodes attempting to authenticate, or false messages being transmitted before authentication has ended.  

\section{Comparison to Existing Designs}
The security holes present in the CAN protocol have led to led researchers to propose a variety of different possible solutions.  These approaches tend to involve adding security features through either changes to the base CAN protocol itself \cite{groza2012libra} \cite{radu2016leia} \cite{aishwarya2016authentication} or proposing frameworks around the existing protocol (such the one we are proposing) so that the CAN protocol itself remains the same \cite{wu2016security} \cite{king2017international} \cite{siddiqui2017secure}.  To the best of our knowledge, there are not many proposed security solutions that explicitly integrate PUFs as a core component of the system.  As such we are not considering systems in which a PUF could replace an existing component such as using a PUF to remove the need for secure nonvolatile memory \cite{feiri2013efficient}.

One example PUF-based solution is the work from \cite{aishwarya2016authentication}.  That work uses PUFs embedded in each ECU to validate the ECU before sending a message to another ECU.  All communication between ECUs is routed through a reference monitor which is responsible for validating the identity of the ECU during each communication before forwarding the associated message to its intended recipient.  We are not considering this work in our comparisons due to the fact that the CAN bus has been effectively replaced by the reference monitor.

A separate work  also uses a PUF and server based approach in which the server and each ECU have an integrated PUF \cite{siddiqui2017secure}. During authentication, the server and each ECU generate ECDH key pairs from responses generated by each PUF. Every ECU generates a ECDH key exchange shared secret with the server and transmits an encrypted copy of its public key.  AES-128 is used for encryption the keys. The public key is transmitted across two CAN frames since AES requires block sizes of 128 bits and the data field in a single CAN frame is only 64 bits.  The server compares the received public keys with the public keys that it stored during an enrollment phase to validate each ECU. The server then sends encrypted copies of each valid public key (2 frames each) to each valid ECU along with a third frame denoting which ECU is associated with that public key.  The ECUs use the received public keys to generate a shared secrets with every other valid node.  The ECUs can then encrypt data being sent to any ECU with a unique key that is only available to the sending and receiving ECUs.  Like before, two CAN frames must be sent for every data transmission to comply with the block size of AES-128. 

There are certain scalability, functionality, and security concerns present in the existing framework that our proposed solution is able to overcome.  The scalability issue lies with the authentication phase.  Each valid ECU must receive encrypted copies of the public keys for all other valid nodes.  For a system with $n$ ECUs, a single ECU will transmit its public key to the server and receive an encrypted copy of each valid public key along with a message indicating the node ID associated with each key.  This means the server must overall transmit $n^2$ public keys and therefore the number of public keys that must be sent will scale quadratically as the number of ECUs increases.  If you consider that each public key sent by an ECU requires 2 CAN frames and each public key transmitted by the server requires 3 frames, then the total number of frames required for authentication is $3n^2$ + $2n$.  In addition, the fact that there is a unique shared secret between every pair of ECUs prevents broadcast messages.  An ECU must separately encrypt and send duplicate messages to each intended ECU.  

As stated in the previous section, our proposed framework will complete authentication after sending $4n$ frames for a system containing $n$ ECUs.  This means the number of required frames scales linearly with the number of ECUs in the system.  The fact that there is a single session key shared by all of the nodes mean that our proposed framework supports messages having multiple intended recipients.  The use of PRESENT for encryption allows an entire encrypted message to fit within the data field of a single CAN frame during normal communication between ECUs.  Table \ref{COMPTAB} shows a comparison between \cite{siddiqui2017secure} and our proposed framework in terms of the total number of CAN frames that must be sent during the different phases of operation within a system containing $n$ ECUs.  The table shows that our approach scales much better for larger systems.  For example, a system with 20 ECUs would require the transmission of 1240 frames under the existing framework while our proposed framework would only require 80.

\begin{table}
\renewcommand{\arraystretch}{1.3}
\caption{Required CAN Frames for $n$ ECU System}
\label{COMPTAB}
\centering
\begin{tabular}{l c c}
\toprule
\bfseries Operation Phase & \bfseries \cite{siddiqui2017secure} 
	& \bfseries Proposed\\
\midrule
Authentication & $3n^2$ + $2n$ & $4n$ \\
Normal Communication & $2$ & $1$\\
\bottomrule
\end{tabular}
\end{table}

This scaling issue becomes even more important when you consider the amount of time it actually takes to transmit a CAN frame.  CAN has both high-speed and low-speed versions.  High-speed CAN can transmit data at speeds of up to 1 Mb/s while low-speed can transmit at speeds of up to 125 kb/s.  Standard CAN frames are 108 bits long.  There is also an extended version which is 128 bits.  Furthermore, CAN requires at least 3 bits of spacing between messages.  This effectively means that standard and extended frames require the transmission of 111 bits and 131 bits, respectively.  Therefore, Low-Speed CAN can transmit a standard and extended frames in 896 $\mu$s and 1048 $\mu$s, respectively.  High-Speed CAN can transmit the frames in 112 $\mu$s and 131 $\mu$s, respectively.

\begin{figure}[h]
\centering
\includegraphics[width=4in]{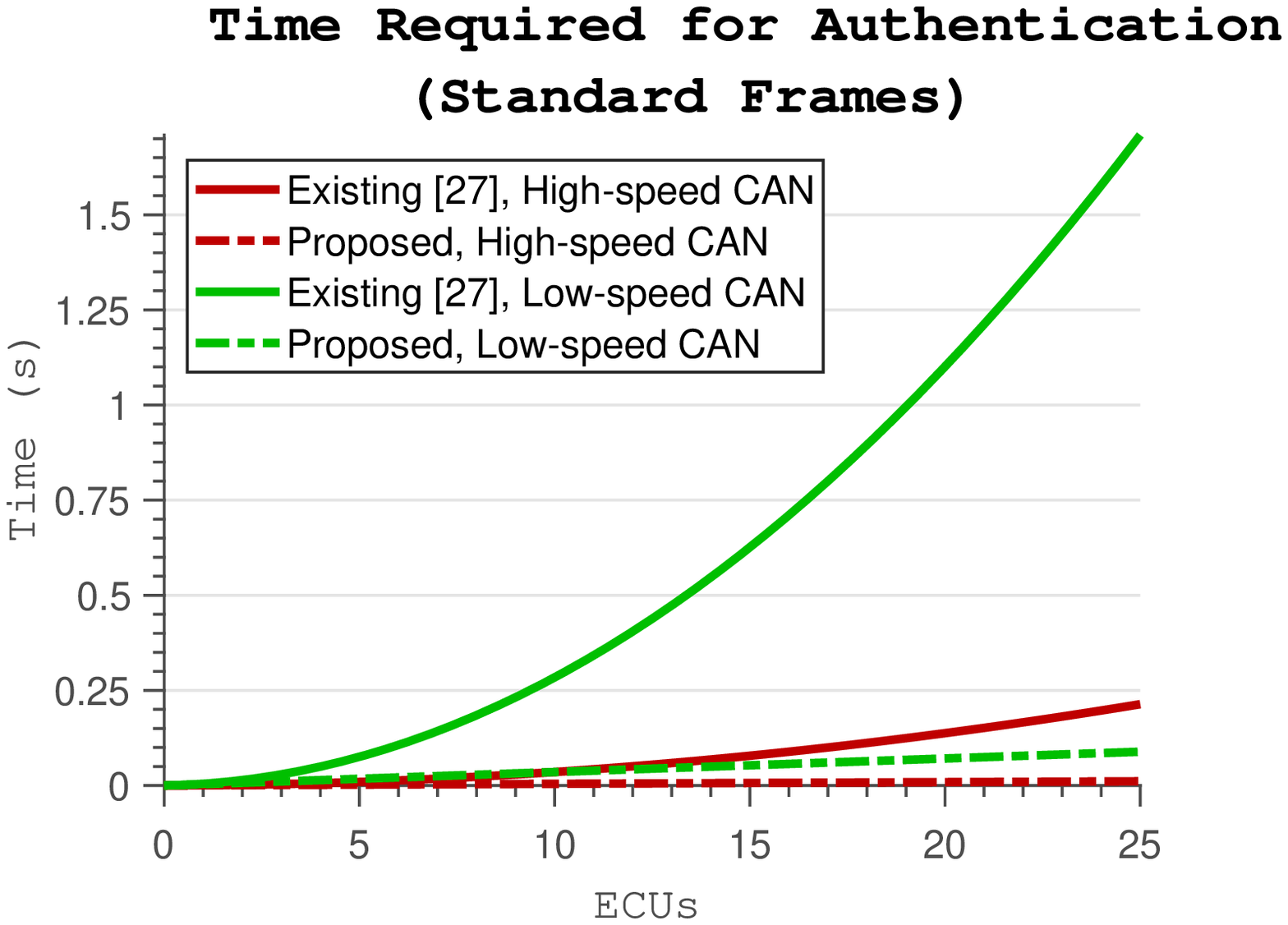}
\caption{Authentication Phase Overhead Comparison (Standard Frames)}
\label{STANDARD}
\Description{Graphical representation of data from Table 2.}
\end{figure}

\begin{figure}[h]
\centering
\includegraphics[width=4in]{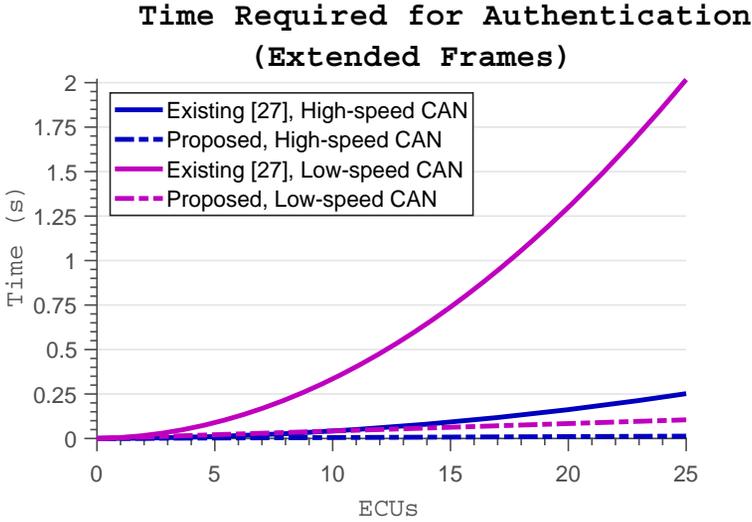}
\caption{Authentication Phase Overhead Comparison (Extended Frames)}
\label{EXTENDED}
\Description{Graphical representation of data from Table 2.}
\end{figure}

Figures \ref{STANDARD} and \ref{EXTENDED} show how long the Authentication would take as the number of nodes increases.  Figure \ref{STANDARD} assumes the system is using standard CAN frames and Figure \ref{EXTENDED} assumes extended CAN frames.  For a system with 20 ECUs, our proposed framework will complete authentication in only 6.5\% of the time that it would take the existing framework.  That percentage will continue to decrease as the number of ECUs increases.  Table \ref{TIMECOMPTAB} contains a comparison of the time required to transmit all of the CAN frames required to complete Authentication with in systems of various sizes.  It is important to highlight the amount of time required for Authentication since it represents extra overhead that is not already present within vehicles.  Implementing these frameworks would require introducing a period of time that the vehicle is unresponsive immediately after it starts.  This period of time might be negligible for systems with a very small number of ECUs, but it will become increasingly pronounced as the number of ECUs increases.  Most importantly, the superior scaling of our proposed framework guarantees that this dead period of operation will remain nearly imperceptible for much larger systems compared to the existing approach.  

\begin{table*}
\scriptsize
\renewcommand{\arraystretch}{1.3}
\caption{Time Required to Transmit the Frames Required for Authentiction}
\label{TIMECOMPTAB}
\centering
\begin{tabular}{l l c c c c c}
\toprule
\bfseries Speed, Frame & \bfseries Framework & \multicolumn{5}{c}{\bfseries Number of ECUs}\\
 & & \bfseries 5 & \bfseries 10 & \bfseries 15 & \bfseries 20 & \bfseries 25\\ 
 \midrule
High, Standard & \cite{siddiqui2017secure} & 9.52~ms & 35.84~ms & 78.96~ms & 138.88~ms & 215.6~ms \\
 & Proposed & 2.24~ms & 4.48~ms & 6.72~ms & 8.96~ms & 11.2~ms\\ \hline
 High, Extended & \cite{siddiqui2017secure} & 11.134~ms & 41.92~ms & 92.36~ms & 162.44~ms & 252.18~ms \\
 & Proposed & 2.62~ms & 5.24~ms & 7.86~ms & 10.48~ms & 13.1~ms \\ \hline
 Low, Standard & \cite{siddiqui2017secure} & 76.16~ms & 286.72~ms & 631.68~ms & 1,111.04~ms & 1,724.8~ms \\
 & Proposed & 17.92~ms & 35.84~ms & 53.76~ms & 71.68~ms & 89.6~ms\\ \hline
 Low, Extended & \cite{siddiqui2017secure} & 89.08~ms & 335.36~ms & 738.84~ms & 1,299.52~ms & 2,017.4~ms\\
 & Proposed & 20.96~ms & 41.92~ms & 62.88~ms & 83.84~ms & 104.8~ms \\
\bottomrule
\end{tabular}
\end{table*}

NIST guidelines state that for symmetric keys of size 128 bits, the elliptic curve key size to provide equivalent security is 256 bits \cite{barker2016nist}.  Per the NIST specifications, security for the 128-bit ECDH key used in \cite{siddiqui2017secure} would actually be equivalent to a symmetric key that is less than 80 bits.  The normal security strength of AES-128 is potentially undercut since the shared secret used as the encryption key is derived from the 128-bit ECDH keys.  This could present a vulnerability that could be exploited by an attacker.  In our approach, the encryption key used during normal operation is a session key that the server randomly generates each time.  During enrollment, we are able to make use of 256-bit ECDH FourQ shared secrets by hashing them to 80-bit or 128-bit keys.  In this way, the security of the encryption function is not reduced by the key generation.

\section{Discussion and Concluding Remarks}
In this paper we present a novel CAN security framework based on PUF.  The proposed framework offers improvements over previous PUF-based frameworks in terms of both scalability and the message overhead associated with normal operation.  The savings in overhead results in our proposed framework being able to send the number of CAN frames required for the Authentication of a system with 20 nodes in only 6.5\% of the time that it takes the existing framework.  Normal message passing in our proposed framework requires only a single CAN frame to be sent while the existing approach requires two frames per message.

Our framework merely uses PRESENT and PHOTON as examples of LWC functions.  PRESENT could be substituted for an alternative with a block size of 64-bits and a key size of at most 128-bits.  PHOTON could be replaced by a different lightweight hash that is capable of producing a 128-bit output.  Ongoing efforts in the development of LWCs, including NIST's efforts to standardize LWC, will likely result in new functions that offer better performance than what is currently available. Depending on the implementation focus, it might be preferable to choose an LWC that was optimized for hardware implementation rather than software implementation or vice-versa.  One interesting avenue for future research would be a comprehensive study on the performance of various LWCs when implemented in both software for various resource-constrained platforms and in hardware such as FPGAs and ASICs.

\begin{acks}
This research was partially supported by grant from National Science Foundation under Grant No:1738662.
\end{acks}

\bibliographystyle{ACM-Reference-Format}
\bibliography{references}

\end{document}